\documentclass[twocolumn,preprintnumbers]{revtex4}
\usepackage{graphicx,bm,amssymb,amsmath,hyperref}

\newcommand{\be}{\begin{equation}}
\newcommand{\ee}{\end{equation}}
\newcommand{\bear}{\begin{eqnarray}}
\newcommand{\eear}{\end{eqnarray}}
\newcommand{\ba}{\begin{array}}
\newcommand{\ea}{\end{array}}
\begin{document}

\preprint{Fermilab-Pub-16-558-T}
\preprint{MITP/16-133}

\title{\boldmath Exotic Signals of Vectorlike Quarks}
\author{Bogdan A. Dobrescu$^*$ and Felix Yu$^\diamond$  \vspace{0.3cm}} 

\affiliation{ $*$ Theoretical Physics Department,  Fermilab, Batavia, IL 60510, USA   \vspace{0.2cm} \\
$\diamond$ PRISMA Cluster of Excellence \& Mainz Institute for Theoretical Physics, Johannes Gutenberg University, 55099 Mainz, Germany \\ } 

\date[]{December 5, 2016; Revised May 5, 2017 \\ \vspace{0.3cm} }

\begin{abstract} \normalsize
Vectorlike fermions are an important target for hadron collider searches. 
We show that the vectorlike quarks may predominantly decay via higher-dimensional operators  into a quark plus a couple of other Standard Model fermions. 
Pair production of vectorlike quarks of charge 2/3 
at the LHC would then lead to a variety of possible final states,
including $t\bar t + 4\tau$,  $t\bar b\nu + 3\tau$, $t\bar t + 4\mu$, $t\bar b\nu + 3\mu$ or $t\bar t + 4b$.
Additional channels ($b\bar b + 4\tau$,  $4t+b\bar b $,  $6b$, etc.) arise in the case of a vectorlike quark of charge $-1/3$. 
If the vectorlike quark decays into three light quarks, then the $3j+3j$ signal is more difficult to observe, 
and  the vectorlike quark mass is almost unconstrained by current searches.
\end{abstract}

\maketitle

\subsection{Introduction \\ [-2mm] }

All elementary fermions discovered so far belong to three generations of chiral quarks and leptons.
Additional  generations of chiral fermions are tightly constrained.
Chiral fermions cannot be heavier than 
about 600 GeV  \cite{Einhorn:1986za} because their masses arise from
couplings to a Higgs  field. 
Searches at the Large Hadron Collider (LHC) set lower mass limits for 
fourth-generation quarks above 1.3 TeV \cite{Khachatryan:2015gza, Aad:2015mba}, assuming decays into a $W$ boson and a
third generation quark. In addition, fourth-generation quarks would increase the Higgs production through gluon fusion by a factor of 9,
while the current measurements allow at most a 40\% increase \cite{Khachatryan:2016vau}. New chiral fermions are also in conflict with the electroweak data \cite{Olive:2016xmw}.
Although Higgs production and the electroweak data constrain loop contributions that can be cancelled by some new particles
running in the loop \cite{Kumar:2012ww}, the existence of a fourth generation of chiral fermions has become very unlikely. 

New elementary fermions  may still exist if they are vectorlike with respect to the Standard Model (SM) gauge group, {\it i.e.},
their left- and right-handed components transform the same way under $SU(3)_c\times SU(2)_W\times U(1)_Y$.
Such vectorlike fermions would represent a new form of matter, and searches for them are a major goal for the LHC experiments.
Pair production of vectorlike quarks at the LHC is large, through the gluon coupling, and depends only on their mass.

By contrast, the vectorlike quark decays are model dependent. The existing searches~\cite{Khachatryan:2015gza, Aad:2015mba} for 
a vectorlike quark of charge $+2/3$ (or $-1/3$)  assume that 
it decays into $Wb$, $Zt$, $ht$ (or $Wt$, $Zb$, $hb$), where $h$ is the SM Higgs boson.
These are the dominant decay modes when the vectorlike quark mixes with a third generation quark \cite{Han:2003wu,Dobrescu:2009vz}. 
In the presence of new bosons, this mixing leads to additional 2-body decays \cite{Serra:2015xfa}.

We point out that the mixing of a vectorlike quark with the SM quarks can be negligibly small, 
and the main decay modes may arise due to effects of additional fields. Such fields do not need to be within the LHC reach,
and their effects may be parametrized by higher-dimensional operators. We focus on effective interactions of a vectorlike quark with 
three SM fermions. The ensuing LHC signals are rather exotic, involving six SM fermions.


\subsection{``Standard" decays of a vectorlike quark \\ [-2mm] }

Let us first consider a vectorlike quark $\chi$ that transforms as $(3, 1, 2/3)$ 
under $SU(3)_c\times SU(2)_W\times U(1)_Y$, {\it i.e.}, $\chi_L$ and $\chi_R$ have the same charges as the SM right-handed up-type quarks $u_R^j$, where $j=1,2,3$ labels the generations.
A mass term $- m_\chi \bar \chi\chi$  is present in the Lagrangian,  with $m_\chi > 0$.  Other gauge invariant mass terms, of the type  $\bar \chi_L u_R^j$, can be rotated away by a redefinition 
of the right-handed fields.
The SM Higgs doublet, $H$, has Yukawa couplings to $\chi$ and to the 
SM quark doublets, $q_L^j$:
\be
- \overline q^j_L   H  \left( y_{j\chi} \chi_R + y_{ji}  u_R^{i} \right)  + {\rm H.c.} 
\ee
For $y_{j\chi} \neq 0$, the $u_L^j$ and $\chi_L$ gauge eigenstates mix. We assume for simplicity that only the 
third generation quarks mix with $\chi$. The mixing angle $\theta_L$ is then
given by
\be
s_L\equiv \sin\theta_L \simeq  y_{3\chi} \frac{v_H}{m_\chi} ~~,
\ee
where we kept only the leading order in $v_H/ m_\chi$, with $v_H \approx 174$ GeV.

The mixing between $\chi$ and the third generation up-type quark leads to two mass eigenstates: the top quark, and a heavier $t'$ quark.
The latter has three decay modes via mixing: $W b$, $Zt$ and $ht$. These
``standard"  decays have a combined width of \cite{Dobrescu:2009vz}
\be  \hspace*{-0.2cm}
\Gamma(t^\prime \!\! \rightarrow W b, Zt, ht) = \frac{s_L^2 (2 - s_L^2)}{32 \pi v_H^2}  m_{t^\prime}^3
\left[1 + O\!\left(\frac{m_t^2}{m_{t^\prime}^2}\right)\right] .
\ee
It is clear that for small mixing the standard $t'$ width may be much smaller than for other, more exotic decays. 
We will focus on models where the exotic $t'$ decays occur at tree level, while the 
$\chi$-$u^3_R$ mixing vanishes at tree level.


\subsection{Exotic decays of a vectorlike quark \\ [-2mm]}

Additional interactions of $\chi$ with SM particles may be mediated by some new
particles that are too heavy to be observable at the LHC. In that case it is convenient to integrate out the heavy particles, 
which generates higher-dimensional operators that include only $\chi$ and SM fields.
An example is a 4-fermion operator that involves one $\chi$, one top quark, and two third-generation leptons: 
\be
\frac{\lambda_\chi \lambda_q}{M_\xi^2}  (\overline \chi_R l_L^3) 
 \, i\sigma_2 (\overline \tau_R q_L^3 )  + {\rm H.c.} 
\label{eq:4ftaus}
\ee
Here  $l_L^3 \equiv (\nu_\tau, \tau)_L$ is the lepton doublet,
$\sigma_2$ is a Pauli matrix acting on the implicit $SU(2)_W$ indices, 
$\lambda_\chi$ and $\lambda_q$ are dimensionless real parameters, and 
$M_\xi$ is a mass parameter.
The operator (\ref{eq:4ftaus}) arises, for instance, from a renormalizable theory that includes a scalar leptoquark $\xi$ transforming as $(3, 2, 7/6)$ under the SM gauge group. 
This is one of the typical sets of quantum numbers considered in leptoquark searches \cite{Olive:2016xmw}. 
The Yukawa  interactions of $\xi$ are $ \lambda_\chi\, (\overline \chi_R    l_L^3) 
i \sigma_2 \xi $ and $  \lambda_q \, \xi^\dagger  (\overline \tau_R q_L^3 ) $, and the $\xi$ mass is 
$M_\xi$.

An additional  Yukawa  interaction allowed by gauge invariance is $ \lambda_t\,  \xi^\dagger i \sigma_2 (\overline l_L^3 u_R^3  ) 
$. 
The exchange of $\xi$ then also induces  
the 4-fermion operator $\lambda_\chi \lambda_t /M_\xi^2  \,  (\overline \chi_R l_L^3) (\overline l_L^3  u_R^3 ) $,
whose effects cannot be neglected unless $\lambda_t \ll \lambda_q$.

The decay $t' \to \tau^+\tau^-t$, mediated by the up-type component of $\xi$, has a width  
\be  \hspace*{-0.2cm}
\Gamma(t^\prime \!\! \rightarrow \tau^+\tau^-t) = \frac{ \lambda_\chi^2 (\lambda_q^2 + \lambda_t^2) }{6144 \pi^3 M_\xi^4}  m_{t^\prime}^5
\left[1 + O\!\left(\frac{m_t^2}{m_{t^\prime}^2}\right)\right] .
\label{eq:taus}
\ee
The down-type component of $\xi$ mediates additional decays: $t' \to \tau^+\nu b$  and $t' \to \nu \bar\nu t$, 
with partial  widths proportional to  $\lambda_\chi^2 \lambda_q^2 $ and $\lambda_\chi^2 \lambda_t^2 $, respectively. The sum of their widths is equal 
to $\Gamma(t^\prime \! \rightarrow \tau^+\tau^-t)$
up to $m_t^2/m_{t^\prime}^2$ corrections.

\begin{figure}[t]
\hspace*{-3mm}\includegraphics[width=0.28\textwidth]{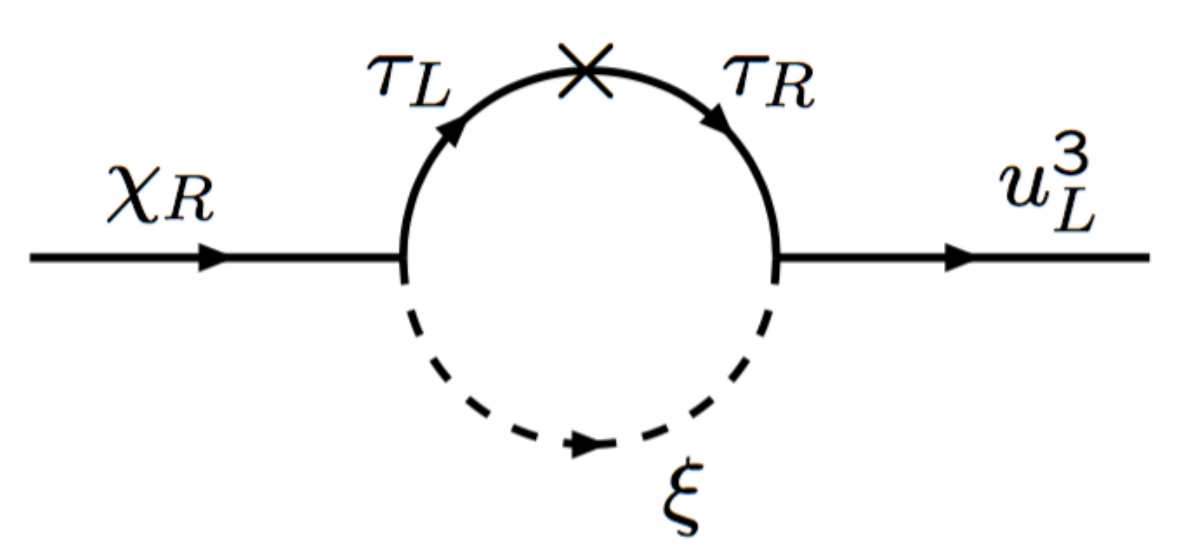}\vspace*{-3mm}
\caption{Loop-induced mixing between the vectorlike quark $\chi$ and the SM quark $u^3$. A tau mass insertion is denoted by $\times$.}
\label{fig:loop}
\end{figure}

Although the limit of vanishing mixing, $s_L \to 0$, can be enforced by a symmetry (for example a $Z_2$ acting on $\chi$) in the Yukawa couplings of $\chi$ and $H$, 
the 4-fermion operator (\ref{eq:4ftaus})  breaks that symmetry. As a result, a nonzero $s_L$ is generated  
at one loop, as shown in Fig.~\ref{fig:loop} when the effective operator is mediated by $\xi$. That contribution is logarithmically divergent:
$s_L =  y_\tau \lambda_\chi \lambda_q v_H /(8 \pi^2 m_\chi) \ln (\Lambda/M_\xi)$,
where $\Lambda$ is  an ultraviolet cutoff (the scale where the $\bar \chi_R H q_R$ coupling vanishes) and $y_\tau$ is the SM $\tau$ Yukawa coupling.
For Yukawa couplings $\lambda_\chi$, $\lambda_q$ of order one and $\Lambda/M_\xi = O(10^2)$, we obtain the 1-loop mixing $s_L \approx 6\times 10^{-4} v_H / m_\chi$.
An additional 1-loop contribution to the mixing is proportional to $\lambda_\chi \lambda_t$.

The radiative decays, $t' \to t g$ and $t' \to t \gamma$, are induced by 1-loop diagrams like the one shown in Fig.~\ref{fig:loop}, with a gluon attached to the 
$\xi$ internal line, or with a photon attached to any internal line. Their amplitudes are not logarithmically enhanced after including diagrams with initial and final 
state radiation~\cite{Deshpande:1981zq}.  The resulting partial width for $t' \to t g$ is given by $\alpha_s / (2\pi) = 2\times 10^{-2}$ times $\Gamma(t' \to \tau^+ \tau^- t)$, 
and hence we can neglect it. The partial width for $t' \to t \gamma$ is further suppressed by a factor of the order of $\alpha/\alpha_s$.

\begin{figure}[tb!]
\hspace*{-3mm}\includegraphics[width=0.47\textwidth]{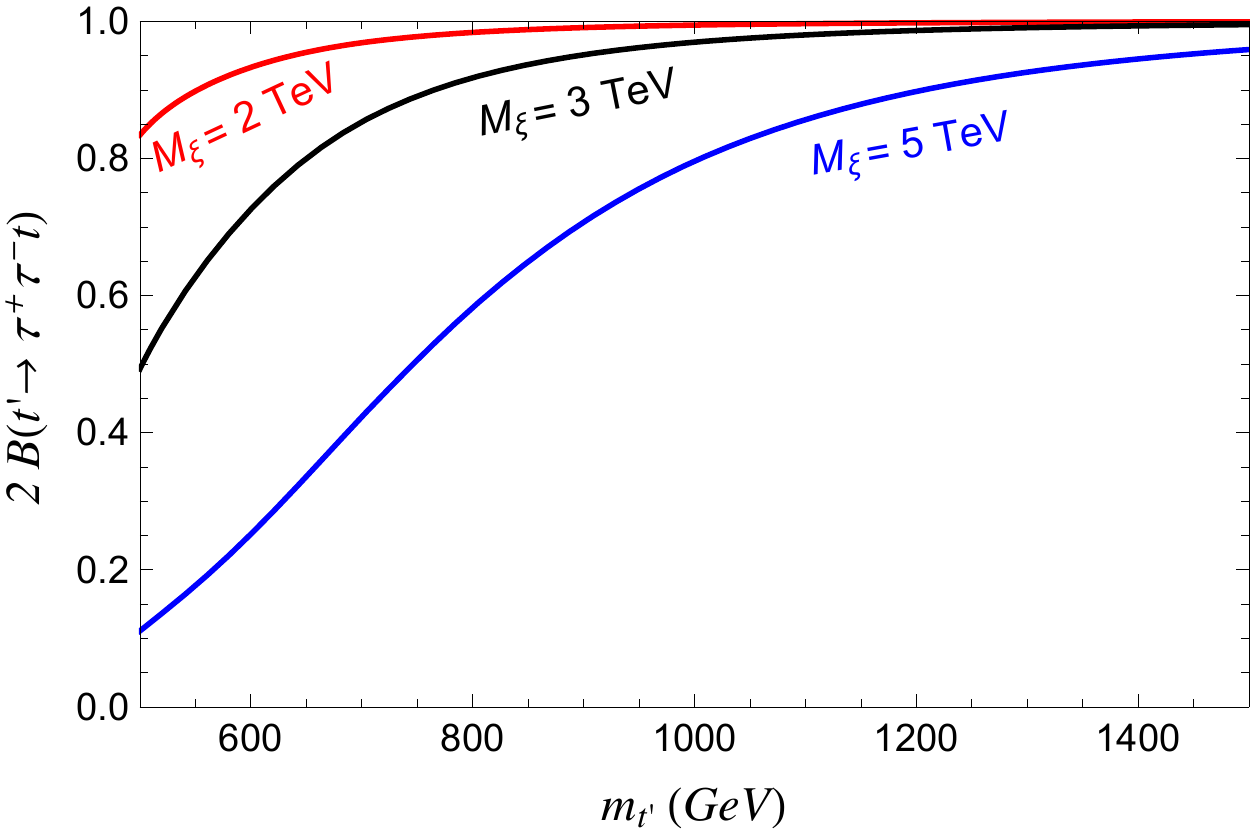}\vspace*{-3mm}
\caption{ Branching fraction of exotic decays  for the 
$t'$ quark, \\ $B(t' \!\to \tau\tau t)+B(t'\! \to \tau\nu b) \approx 2 B(t' \!\to\tau^+\tau^- t)$,
 when the $s_L$ mixing is induced at one loop and $\lambda_t \to 0$. 
The mass of leptoquark $\xi$ is fixed at 2, 3 and 5 TeV. }
\label{fig:Btp}
\end{figure}

The exotic decays have a branching fraction above 80\% for
$m_{t^\prime} \! \gtrsim M_\xi/4$, as shown in Fig.~\ref{fig:Btp} for
$\lambda_t \ll \lambda_q$.  The $m_{t'}/M_\xi$ corrections are
included there using MG5\_aMC \cite{Alwall:2014hca} with model files
generated by FeynRules \cite{Alloul:2013bka}.  We remark that these
exotic decays share the same final state as the subdominant decays in
standard channels, $t^\prime \to t h, h \to \tau^+ \tau^-$ and
$t^\prime \to t Z, Z \to \tau^+ \tau^-$, but the enhanced exotic
branching fraction effectively reweights the sensitivities in
dedicated searches~\cite{Khachatryan:2015gza, Aad:2015mba}.  In
particular, the relatively high leptonic branching fraction of $\tau$s
enhances the sensitivity of dedicated multilepton analyses for
vector-like quarks, while weakening the constraint from traditional $h
\to b\bar{b}$ or $Z \to \ell^+ \ell^-$ search subchannels.  In
addition, the $\tau$ kinematics from the operator (\ref{eq:4ftaus})
will be notably different from those predicted in the overlapping
standard decays.  In particular, the sequential 2-body decays become
more energetic as the $t^\prime$ mass increases, leading to $\tau$
pairs that become more boosted for heavier $t^\prime$.  On the other
hand, the $\tau$ particles from the exotic 3-body decay remain
isotropic, which will prove an important handle in post-discovery
characterization.

Four-fermion operators similar to (\ref{eq:4ftaus}), with the taus replaced by muons or electrons, are also possible,
and lead to LHC signatures with smaller backgrounds. The 1-loop value of the mixing, which is proportional to the lepton mass, is much smaller in that case. As a result,
the exotic $t'$ branching fraction remains large for heavier $\xi$. 
For example, if the leptoquark couples only to second generation leptons (which can be enforced by a 
lepton flavor symmetry), then the operator
$ (\overline \chi_R l_L^2) i\sigma_2 (\overline \mu_R q_L^3) $ leads to $t' \!\to \mu^+\mu^- t$ and $\mu^+\nu b$ decays with a 
combined branching fraction of 98\% when $M_\xi = 10$ TeV and $m_{t'}= 1$ TeV. 

If the $t'$ quark decays predominantly through the 4-fermion operators mediated by $\xi$,  its decay length is
\be
L_{t'} = \frac{1.9  \; {\rm  \mu m} }{\lambda_\chi^2 (\lambda_q^2 + \lambda_t^2) } \left( \frac{M_\xi}{100 \; {\rm TeV}} \right)^{\! 4}
 \left( \frac{ 1 \; {\rm TeV}} {m_{t'}} \right)^{\! 5}   ~~.
\ee
For a leptoquark of mass below  100 TeV  and Yukawa couplings of order one, the decay length of a $t'$ of mass near the TeV scale is 
shorter than 1 $\mu$m. Thus, $t'$ has prompt decays even for a $\xi$ heavier by two orders of magnitude than the LHC reach.
The case where the exotic vectorlike decays produce displaced vertices ($M_\xi /\lambda_\chi\gtrsim 300$ TeV) is also interesting, but will not be discussed here.


\subsection{Four-quark operators \\ [-2mm] }

Other mediators may generate 4-quark operators involving one $\chi$ and three SM quarks. Consider, for example, the operator 
\be
\frac{\kappa_\chi \kappa_t}{M_\zeta^2}  (\overline \chi_R^c  d_R^3)(\overline{d}_R^3 u^{3 c}_R)  + {\rm H.c.} \ ,
\label{eq:4fud}
\ee
where all the quark fields are written in the gauge eigenstate basis, the index $c$ denotes charge conjugation, and
the implicit color indices are contracted through antisymmetric tensors.
This operator arises from the exchange of a scalar diquark $\zeta$, whose gauge charges are $(3, 1, -1/3)$, and whose 
mass is $M_\zeta$.
Its Yukawa interactions 
are given by $\kappa_\chi  \, \zeta \, \overline \chi_R^c  d_R^3$ and $\kappa_t  \, \zeta^\dagger \, \overline{d}_R^3 u^{3 c}_R$.

The ensuing $t' \to b \bar b t $ decay has a width
 \be  \hspace*{-0.2cm}
\Gamma(t^\prime \!\! \rightarrow b \bar b t ) = \frac{ (\kappa_\chi  \kappa_t)^2 }{2048 \pi^3 M_\zeta^4} \,  m_{t^\prime}^5
\left[1 + O\!\left(\frac{m_t^2}{m_{t^\prime}^2}\right)\right] .
\ee
Even though the mixing induced by a $b$ quark loop is larger than that due to a $\tau$ loop by a  $y_b / y_\tau$ factor ($y_b$ is 
the $b$ Yukawa coupling),  the exotic branching fraction is large for a range of parameters; {\it e.g.},  for  $m_{t'} = 1$ TeV
and $M_\zeta$ increasing from  2 TeV and 5 TeV, $B(t^\prime \!\! \rightarrow b \bar b t )$ decreases from 98\% to 50\%.

The operator $ (\overline \chi_R^c  u_R^3)(\overline{u}_R^3 u^{3 c}_R)  $,   induced by a scalar diquark of charge $-4/3$,
would lead to the $t' \to t t \bar t$ decay.  However, the 1-loop mixing in this case is proportional to the large top Yukawa coupling, so that the $t' \to t t \bar t$
branching fraction is smaller than the ``standard" one.

Another operator, also induced by the exchange of 
a scalar diquark carrying  gauge charges $(\bar{6}, 1, -4/3)$, is
\be
\frac{\kappa_\chi' \kappa_t'}{M_{\zeta^\prime}^2}  (\overline \chi_R^c  u_R^2)(\overline{u}_R^2 u^{2 c}_R)  + {\rm H.c.} 
\label{eq:4fcharm}
\ee
Note that the gauge eigenstate $u_R^2$ may be identical with the right-handed charm quark in the mass eigenstate basis.
In that case, there are no worrisome flavor-changing processes. The decay induced by the above operator, $t' \to cc\bar c$,
is difficult to observe due to large backgrounds at the LHC. 
The width of this decay is analogous to that for $b\bar b t$, with  $\kappa_\chi$, $\kappa_t$ and $M_\zeta$ replaced 
by the corresponding primed parameters. The standard decays, though, are more suppressed in this case because the 1-loop $s_L$ is proportional to 
the charm Yukawa coupling instead of $y_b$.  For example,   $m_{t'} = 0.5$ TeV
and $M_{\zeta'}$ in the 1--2 TeV gives $B(t^\prime \!\! \rightarrow c \bar c c )$ in the 99.6\%--93.4\% range. Note that the large exotic branching fractions as well as the nonresonant kinematics in the outgoing light flavor jets give very different experimental handles compared to the standard hadronic channels of vectorlike quark searches.


\subsection{Other vectorlike quarks \\ [-2mm]}

So far we have only studied the vectorlike quark that carries electric charge 2/3 and is an $SU(2)_W$ singlet.
A similar discussion applies to any other vectorlike fermion, albeit the collider signals are different in some cases.
Let us briefly discuss a vectorlike quark $\omega$ that transforms as $(3, 1, -1/3)$, and does not mix with the SM quarks at tree level.
The operators 
\be
\frac{\lambda'_\chi }{M_{\xi'}^2}  (\overline \omega_R \tau_R^c ) \left(  \lambda'_q \, \overline l_L^{3 c } i\sigma_2  q_L^3 +   \lambda'_t \,  \overline \tau_R^c  d_R^3 \right)  + {\rm H.c.} 
\label{eq:4fomega}
\ee
are induced by a scalar leptoquark $\xi'$ of SM charges $(3,1, -4/3)$. 
At one loop, this operator leads to mixing between $\omega$ and the SM $b$ quark proportional to $y_\tau$. We label the heavy quark mass eigenstate by $b'$. The decay 
$b' \to b \tau^+\tau^-$ has a width similar to the one in Eq.~(\ref{eq:taus}), with  $\lambda_\chi$, $\lambda_q$, $\lambda_t$, $M_{\xi}$ replaced by the primed parameters.
The width for $b' \to t \tau^- \nu$ is proportional to $(\lambda'_\chi \lambda'_q)^2$, and the $b' \to b \nu \bar \nu$ process is loop suppressed.

Four-quark operators may lead to $b'\to bb\bar b$ or  $b'\to bc\bar c$ with large branching fractions.
Vectorlike quarks carrying charge 5/3 or $-4/3$ may also have exotic decays involving two leptons and one quark, or three quarks.

\medskip

\begin{figure}[tb!]
\hspace*{-3mm}\includegraphics[width=0.46\textwidth]{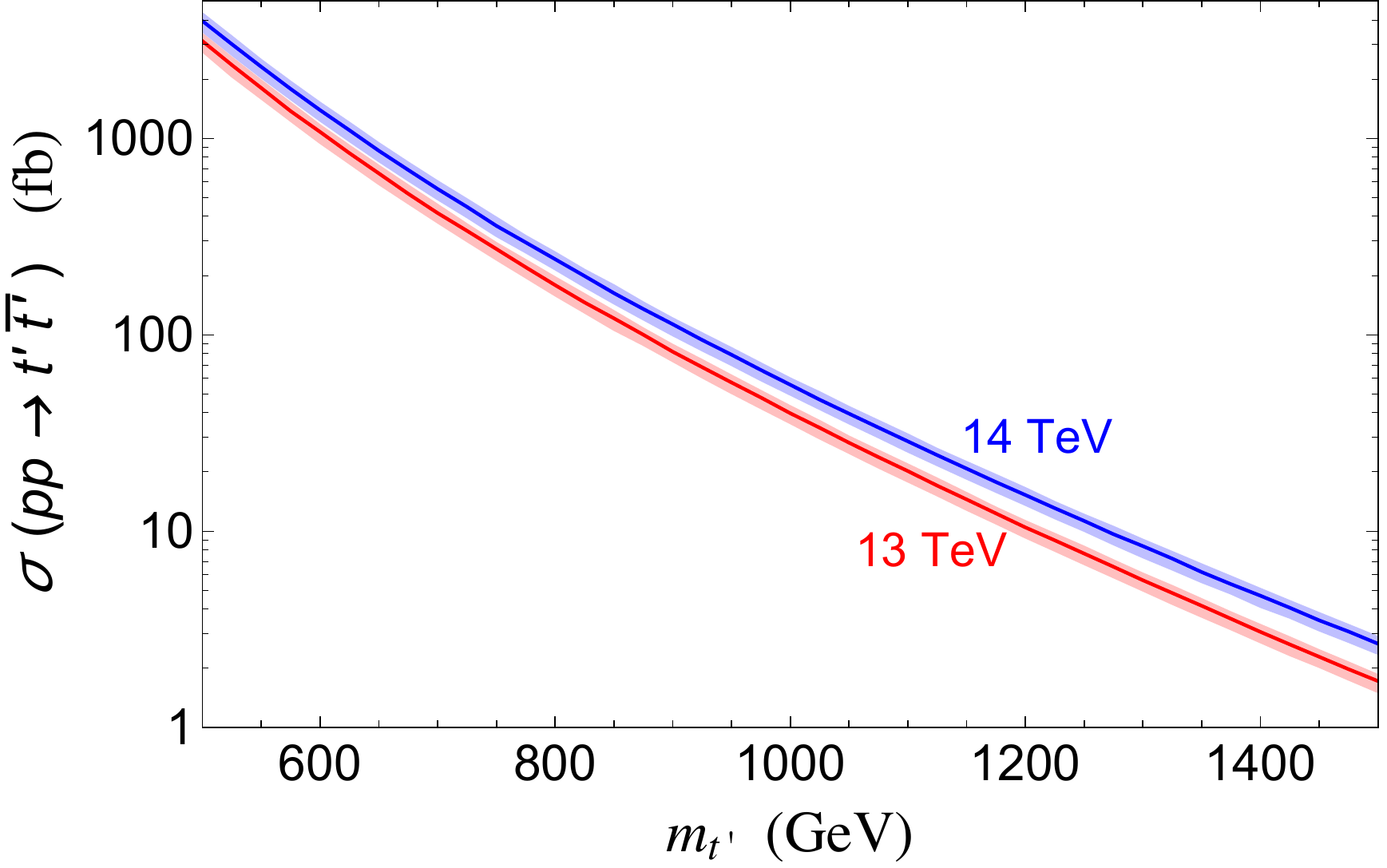}\vspace*{-3mm}
\caption{ Next-to-leading order cross sections for vectorlike quark pair production at the 13~TeV (red) and 14~TeV (blue) LHC, calculated with MCFM v8.0.} 
\label{fig:LHCxsecs}
\hspace*{-3mm}\includegraphics[width=0.33\textwidth]{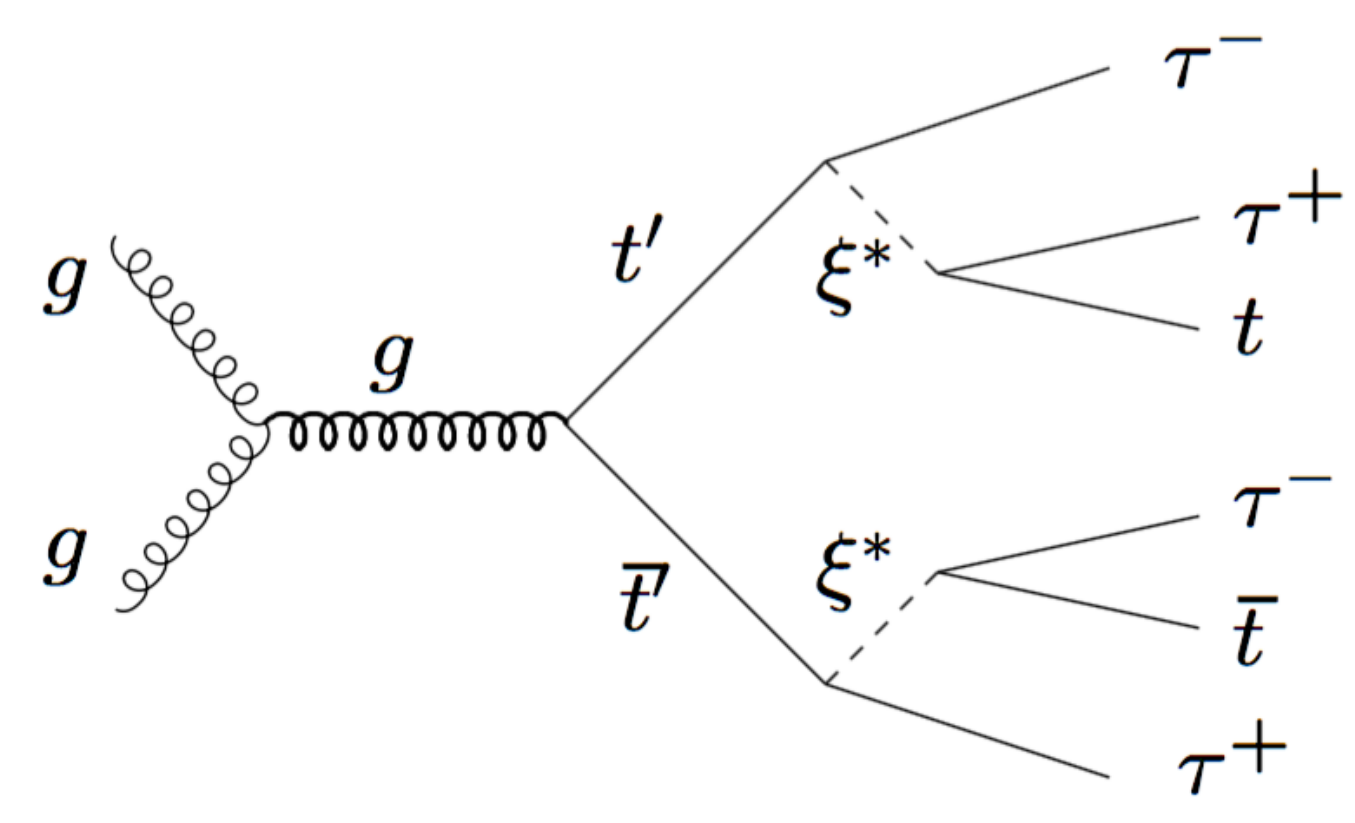}\vspace*{-3mm}
\caption{ Representative diagram for pair production of vectorlike quarks at the LHC followed by exotic decays into taus and top quarks via off-shell leptoquarks $\xi$. \\ [-3mm] }
\label{fig:TauTop}
\end{figure}

\subsection{LHC signals of vectorlike quark pairs \\ [-2mm] }

At the LHC, pair production of vectorlike quarks proceeds readily from strong interactions, leading to large cross sections ranging from 
picobarns for vectorlike quark masses around 500 GeV to femtobarns for masses around 1.5 TeV.  We show the pair-production cross section at next-to-leading order in $\alpha_s$ for color-triplet vectorlike quarks in Fig.~\ref{fig:LHCxsecs}, which was calculated using MCFM v8.0~\cite{Campbell:1999ah} with CT14 parton distribution functions~\cite{Dulat:2015mca} evaluated at renormalization and factorization scales equal to the vectorlike quark mass.  The shaded bands show the cross section dependence after varying these scales within the $m_{t'}/2 - 2m_{t'}$ interval.
Single production of the vectorlike quark is suppressed by $s_L^2$, and can be ignored here.

The 3-body decays of the vectorlike quarks lead to six SM fermions in the final state. In Fig.~\ref{fig:TauTop} we show a diagram for pair production of vectorlike 
quarks leading to the $t \bar t +4\tau$ final state, via off-shell leptoquarks. This high-multiplicity of third generation SM fermions gives various same-sign lepton, 
multi-lepton, and hadronic $\tau$ signatures, each accompanied by $b$-tagged jets.   The total branching fraction for this process, $B(t^\prime \! \rightarrow \tau^+\tau^-t)^2$, 
can be nearly 1/4 (see Fig.~\ref{fig:Btp}).  Focusing on this process, we recast the existing single lepton, dilepton, and multilepton analyses from the 8~TeV analyses~\cite{Khachatryan:2015gza} to constrain the $t'$ decaying via the rates prescribed in Fig.~\ref{fig:Btp}.  We find the multilepton channels of these searches to be the most powerful, since the large multiplicity of $\tau$s and their relatively large leptonic branching decays give high efficiencies for the multilepton selection cuts.  The corresponding existing bounds from these analyses is 650~GeV.  The main SM bankgrounds in the multilepton channel include $WZ$ and $WWW$ production, as well as $t\bar{t}$ production with misidentification of a lepton.  Data control samples are used in Ref.~\cite{Khachatryan:2015gza} to determine these backgrounds, resulting in a SM cross section times efficiencies of order 0.2~fb at the 8~TeV LHC.

Other final states induced by the same leptoquark exchange are $t b \nu +3\tau$,  $t\bar t \tau^+\tau^- \nu\nu$, $t b \tau +3\nu$ or $t\bar t + 4 \nu$, which feature 
$b$ jets, missing transverse energy and other kinematic handles.
For large vectorlike masses, the total invariant mass of the observed jets is a strong discriminant against the SM backgrounds.  A large $m_{t'}$ 
also gives significant boosts to the decay products, leading to patterns in jet substructure reconstruction.

If the main decay mode of $t'$ is mediated by operator~(\ref{eq:4fud}), then the final state at the LHC is $t\bar t +4 b$. The presence of six $b$ jets in the final state 
allows a good discrimination of the background, although keeping a large enough signal would require that only some of the jets are tagged.  While each $t'$ is in principle fully reconstructible, we expect the intrinsic combinatorial jet background will be a significant obstacle to reconstruct the $tb\bar b$ resonance.  

Pair production of $b'$ quarks leads to an additional final state, $b\bar b+4\tau$, as well as to some of the final states obtained through  $t'\bar t'$ production ($t\bar t \tau^+\tau^- \nu \bar\nu$, $t b \tau + 3\nu$).  
Useful signals include three leptons, or two leptons and a hadronic $\tau$.  The backgrounds are again due to the production of two or three weak bosons, and from mistidentification of a lepton or a hadronic $\tau$ in other processes; the latter can be obtained from data control samples~\cite{Khachatryan:2015gza}.
If the operators involve muons or electrons instead of $\tau$'s, then the final states have smaller backgrounds and can be more easily reconstructed.

\begin{figure}[tb!]
\hspace*{-3mm}\includegraphics[width=0.48\textwidth]{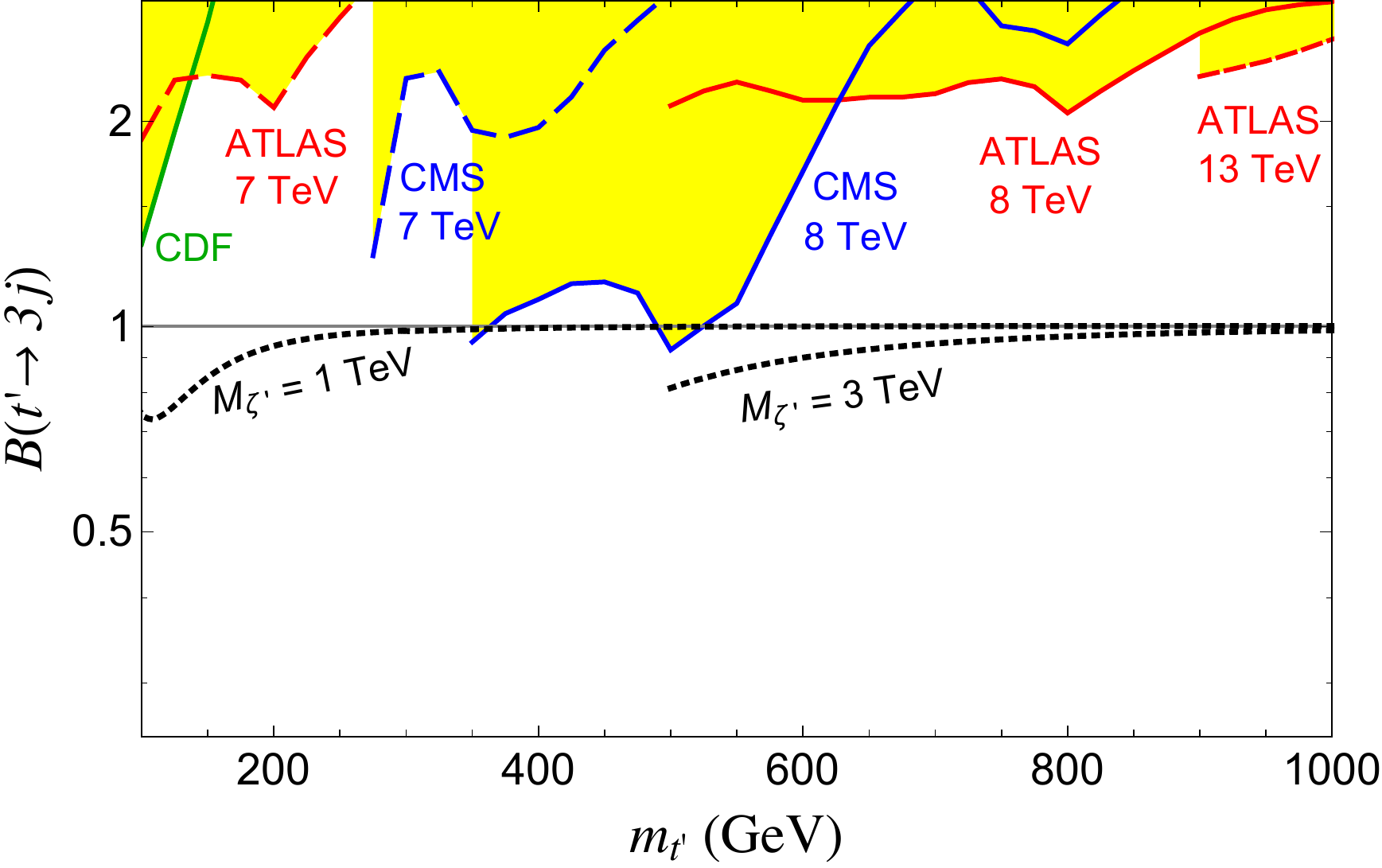}\vspace*{-2mm}
\caption{Limits on the 3-jet branching fraction of a color-triplet fermion produced in pairs
as a function of its mass. Shaded regions are ruled out at the 95\% confidence level by 
searches of the $3j+3j$ signal  at the LHC and the Tevatron. The physical region, $B(t' \!\to 3j) \leq 1$,
remains unconstrained except for a tiny region around $m_{t'} = 500$ GeV. The dotted lines are the predicted $B(t' \!\to cc\bar c)$ in the model associated with operator (\ref{eq:4fcharm}), for $M_{\zeta'} = 1$ TeV and 3 TeV. }
\label{fig:B3j}
\end{figure}

If the dominant decay mode of a vectorlike quark is into three jets, as in the case of operator (\ref{eq:4fcharm}), then the 
signal at hadron colliders is a $3j+3j$ final state, with each set of three jets reconstructing a resonance of equal mass.
This signal was searched for  by CDF \cite{Aaltonen:2011sg} at the Tevatron, and by CMS and ATLAS at the LHC using 
data from the 7 TeV \cite{ATLAS:2012dp},  8 TeV  \cite{Chatrchyan:2013gia}
 and 13 TeV \cite{Aaboud:2018lpl} runs. 
Those results are presented as mass limits on the gluino, which is a color-octet fermion. We adapted those limits 
to our case of color-triplet fermions, whose pair production is reduced due to a smaller Casimir factor. We present the 
limits on vectorlike quarks in Fig.~\ref{fig:B3j}, in terms of the 3-jet branching fraction $B(t' \!\to 3j)$ as a function of mass.
For most values of the mass, the existing searches do not constrain the physical region $B(t' \!\to 3j) \leq 1$. 
Nevertheless, further searches with larger data sets at the 13 TeV LHC will likely be sensitive to the $3j+3j$ signal 
for a wide range of masses.

\medskip

\subsection{Conclusions \\ [-2mm] }

We have identified many signatures of vectorlike quarks that are different than those assumed in collider searches thus far.
If the mixing $s_L$ between the vectorlike quark and the SM quarks vanishes at tree level, then the dominant decays of the vectorlike quark may 
be into three SM fermions. The new particles that mediate these decays may be too heavy 
to be directly observed at the LHC.
The exotic decays of vectorlike quarks lead to signatures at the LHC that involve two quarks (often including the top) and four leptons (especially $\tau$'s),
or six quarks.
The multiplicity and kinematics of quarks and leptons in these processes provide a handle for characterizing the signatures.  

We have not discussed the possibility that the mediators leading to the 4-fermion operators in (\ref{eq:4ftaus}), (\ref{eq:4fud}), or (\ref{eq:4fcharm}) are on-shell.  In that case, the exotic decays proceed as sequential 2-body cascade decays and can clearly dominate over the ``standard'' decays (related processes are studied in \cite{Anandakrishnan:2015yfa}).  Moreover, the main production mode for the mediators at the LHC may be through such cascade decays.

If the dominant decay of a vectorlike quark is into three jets, then the current mass limits may be substantially loosened due to the large backgrounds 
in the $3j+3j$ final state.
Mixed topologies involving vectorlike quark pair production followed by an exotic decay and a standard decay may provide more sensitive channels. 

The simple, renormalizable models presented here readily motivate various searches for final states with six SM fermions at the LHC and beyond. 
Dedicated searches by the ATLAS and CMS collaborations in these channels would provide new probes for the existence of vectorlike fermions.

\smallskip  \smallskip


\noindent
{\it Acknowledgments:} 
FY is supported by the Cluster of Excellence Precision
Physics, Fundamental Interactions and Structure of Matter (PRISMA-EXC
1098), the ERC Advanced Grant EFT4LHC of the European Research
Council, and the Mainz Institute for Theoretical Physics (MITP).
BD is grateful to the MITP for hospitality and partial support during the completion of this work. 


\end{document}